# Thinging for Software Engineers

Sabah S. Al-Fedaghi

Computer Engineering Department
Kuwait University
Kuwait
sabah.alfedaghi@ku.edu.kw

*Abstract*—The aim of this paper is to promote the terms *thing* and *thinging* (which refers to the act of defining a boundary around some portion of reality and labeling it with a name) as valued notions that play an important role in software engineering modeling. Additionally, we attempt to furnish operational definitions for terms *thing*, *object*, *process*, and *thinging*. The substantive discussion is based on the conception of an (abstract) machine, named the Thinging Machine (TM), used in several research works. The TM creates, processes, receives, releases, and transfers things. Accordingly, a diagrammatic representation of the TM is used to model reality. In the discussion section, this paper clarifies interesting issues related to conceptual modeling in software engineering. The substance of this paper and its conclusion suggest that thinging should be more meaningfully emphasized as a valuable research and teaching topic, at least in the requirement analysis phase of the software development cycle.

## I. INTRODUCTION

The current norm in software engineering is the object model, in which object orientation has become the standard for the analysis and design phases of the software development process. This model "in object-oriented analysis and design provides a more realistic representation, which an end user can more readily understand" [1]. The model has assimilated ontological issues that explicitly specify the conceptualization of the domain of concern, for which the term *object* represents a fundamental notion in the object-orientation paradigm. This paper is oriented toward modeling the domain of interest with *things*, a notion that is more general than that of objects. *Thing* is interchangeable with *entity* and is applicable to any item that is acknowledged by a system, whether that item be particular, universal, abstract, or concrete [2].

### A. Specific Aim of This Paper

Several papers submitted to software engineering journals and conferences have advanced objections to the use of the term *thing* as "a vague and empty word [that lacks] any definition." One purpose of writing this paper is to defend this term and demonstrate that *thing* specifically and *thinging* in general are as "celebrated" [3] as the terms *object* and *class*. *Thinging* refers to "defining a boundary around some portion of reality separating it from everything else and then labeling that portion of reality with a name" [4]. According to Heidegger, to understand the thingness of things, one needs to reflect on the power of things to "gather" space and time [5]. Thinging expresses how a "thing things", which he explained as "gathering", uniting, or tying together its constituents. Uniting here can be illustrated by the bridge that makes the environment (banks, stream, and landscape) into a unified whole.

According to Fry [6], "The thingly character of the thing does not consist in its being a represented object, nor can it be defined in any way in terms of the objectness, the over-againstness, of the object." "Things" are irreducible to "objects" [7], and the two notions are "incommensurable" [8].

The notions of thing and thinging play an important role in modeling contending with the salience of the widely acclaimed significance of the word *object*, the term currently in vogue among most software engineers.

In computer science, interest in things and thing-orientation [9] dates back to ThingLab (1979) and Self (1987), the programming languages. More recently, Water, a prototype-based language, has linked every XML tag with its top-level ancestor, a "Thing". Imbusch et al. [9] noted that "Thing-oriented programming is the art of creating software composed of Things."

This article is about modeling thinging. Additionally, this paper unpacks philosophical issues that inform the world of computing.

Philosophers attempt to find the essential or deeper meanings of . . . words that refer to important concepts that we use to guide us in making important decisions . . . [and] to a large extent, is to organize these meanings into coherent frameworks that help us make sense out of the world around us. [10]

That being said, a similar value is attached to the potential insights from recognizing the capacity of computers and information technology to shed new light on philosophical issues and pose questions that cannot readily be approached within traditional philosophical frameworks [11].

Specifically, this paper discusses the ontological status of objects and related notions, such as processes and events. Many research works use the term *entity*, but "there is little, in the texts, to differentiate between entities and objects" [12]. Most of the time, an entity is defined in terms of a thing (e.g., in Chen's [13] description of an entity, it is a thing that can be distinctly identified).

The substantive discussion is based on the conception of an (abstract) machine (an assemblage) named the Thinging Machine (TM), which has been used in several research works [14-23]. The main motivation is to justify adopting a terminology that relies more on the notion of things than it does on objects, particularly in the context of the TM.

### B. What Is an Object?

In the object-orientation literature, an object is described in terms of having an identity, state, behavior, and properties, as well as a specified set of operations. Objects, here, include virtual objects (e.g., a web page), ordinary physical objects, (e.g., a building), and institutional entities [24] (e.g., the act of buying). "The world being modelled is made up of objects . . . objects are just there for the picking!" [25]. "Identifying objects is pretty easy to do. Start out by focusing on the problem at hand and ask yourself 'what are the *things* [italics mine in this problem?'" [26]. Objects have a dual nature that turns on their two sets of properties: functional properties and structural properties [27]. Functional properties are related to what an object does (e.g., a car is used for transportation) and its structural properties pertain to its physical makeup (e.g., the car is red and has white seats) [28].

In object-oriented analysis and design, an object models some *unity* [italics mine] that exists in physical or conceptual space, or some new *unity* [italics mine] that could be realized in the physical space because someone has thought it out. [12]

Many descriptions of objects may oftentimes include examples of relevant things that fit into an object's category [9]. According to Maciaszek [29], "an object is an instance of a 'thing'," and "a generic description of a 'thing' is called a class." All the objects common to everyday life, such as paper clips, tablets, and dog collars, are intentionally produced things [28]. Interestingly, in image analysis studies, an object is defined as "a set of regions located near the center of the image, which has significant color distribution compared with its surrounding (or background) region" [30]. Thus, an empty beach at sunset, with red sky, blue sea, and gray sand, has no object but certainly is a (beautiful) thing.

According to Atkins [31], to objectify a thing is to reduce it, to break it down into increasingly smaller parts instead of taking it holistically as it is. An object consists of its universal form with shared particular qualities (e.g., its color, shape, size, and texture are accidental or unnecessary).

What separates an object from any ordinary "thing" is its phenomena of perception as conjured by a subject. Thus, objects are entities that a subject projects desire and necessity, supporting the theory of objectivity and establishing objecthood. [32]

### C. What Is a Thing?

According to Edwards [33], a thing is surely among the most colorless of English words. Almost anything can be labeled with the word *thing*—a word that seems simultaneously essential and empty—and is essential because of its very emptiness. *Thing* is "a banal term we use for designating what is out there, unquestionably . . . what lies out of any dispute, out of language" [3].

Heidegger [5] distinguished between objects and things: "The handmade jug can be a thing, while the industrially made can of Coke remains an object" [3]. For Heidegger [5], things have unique "thingy Qualities" [3] that are related to reality and therefore not typically found in industrially generated objects. According to Heidegger [5], a thing is self-sustained, self-supporting, or independent—something that stands on its own. The condition of being self-supporting transpires by means of *producing* the thing.

The TM, which is based on the concept of thinging, is an abstract machine that creates, processes, and exchanges things. Although, as noted above, several works have described the TM, the following section provides an interpretation of it from a novel perspective.

## II. THINGING MACHINE

Thinging signifies the following:

- Forming, molding, shaping, and refining the "clay-like stuff" of reality to generate things: diverse pieces have their own identities and different compositions, in terms of parts and wholes. The resultant (so-called mereological) universe consists of conceptualized things that we refer to as components of a system.
- The flow of these things in terms of five stages: creation, processing, receiving, transferring, and releasing. "Not only do things exist in the world, but stuff *happens* to them, There are occurrences. There is movement" [31].

Thinging, from our perspective—which deviates from the Heideggerrian thought—is a thing forming itself in the world as a machine. A *thinging machine* (this term is taken from [34]) generates and handles the thing and its constituent subthings (e.g., an object is a machine and its qualities are submachines). The machine (human and non-human) can craft things. Accordingly, this (abstract) machine is defined in terms of its functions to create, process (change), receive, release, and transfer things, as shown in Fig. 1. Additionally, the TM model utilizes triggering (denoted by a **dashed arrow**) to establish connection with other machines that have different type of things. The machine is the building block of that what things.

A machine that crafts things is itself a thing that is crafted by other machines. For example, a human being *is a machine* that includes sensory and cognitive submachines and so forth; simultaneously, a human being *is a thing* in other larger machines such as social machinery (see Fig. 2). Thus, every "thing" is a machine (environment/place) of thinging other things.

Going by the function of a TM, we define a thing as follows:
*A thing (material and immaterial) is what manifests itself in creation, processing, receiving, releasing, and transferring stages of a thinging machine.*

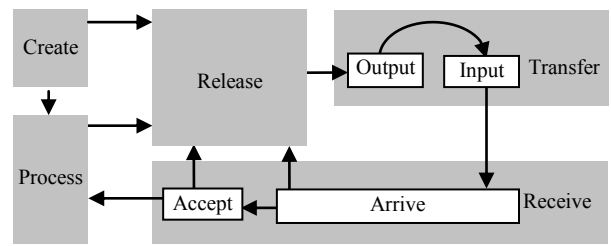

Figure 1. Thinging machine.

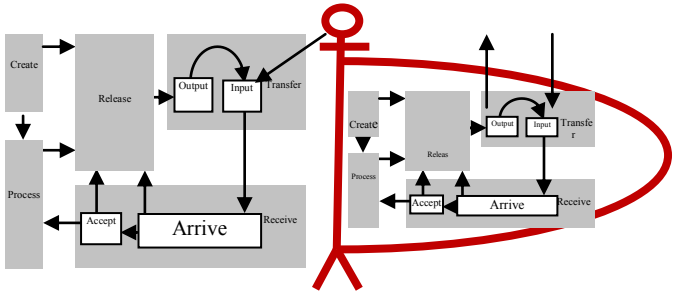

Figure 2. A human being is a thing and a machine.

Examples of things include numbers, time, events, and data. A TM is the "context" of a thing. For example, according to Grigg [35], "The rain, of course, must be raining because it can do nothing else. The 'rain' is its 'raining', just as a 'thing' is its 'thinging.'" The rain is not to be separated from its context and made a static "thing"; rather, it is an operating, dynamic machine (Grigg [35] calls it a *PROCESS*), through which rain is created (in the atmosphere), released, and transferred, to be received by Earth.

Accordingly, instead of "the thing *things*", in a TM, we have five kinds of thingings: the thing *emerges*, *changes*, *arrives*, *transfers*, and *waits* (*for departure*). Thinging is the emergence, changing, arriving, departing, and transferring of things. Heidegger's fourfold concerns the creation type of thing.

Heidegger's notion of thinging has influenced thinking in many scientific fields (e.g., design thinking [36], information services [37], and organization/management studies [38]). The utilization of thinging in this paper is not about the philosophical issues related to the ontology of things and their nature; rather, it concerns the representation of things in software engineering modeling. This representation is utilized in documentation and in the early phases of building software systems.

In several papers submitted to software engineering journals and conferences, some referees rejected the use of the term *thing*. According to one referee, "the system is badly described and many terms are not well defined (e.g., 'thing')." Another referee stated:

"Things can be concepts, actions, or information." This is a very fuzzy explanation. First of all, concepts are independent of space and time, though things are closely related to processes, which are in space and time.

However, the sentence "Things can be concepts, actions, or information" furnishes examples of what can be created, processed, received, released, and transferred. For instance, a *concept* is created, or generated, in the brain; it is processed to create a corresponding proposition; it is received by a listener or reader after being embedded into a speech; it is released in the form of a linguistic expression; and it is transferred from one person to another.

According to Malafouris [39], we are creative "thingers" in the sense that "We make new things that scaffold the ecology of our minds, shape the boundaries of our thinking and form new ways to engage and make sense of the world." The aforementioned referees' comments show little appreciation for thinging and the issue of "defining boundaries around portions of reality" [4] or a significant disregard to the difficulty of defining the problem of *what a thing is*. This is an important aspect to consider with regard to the TM. Lacking a clear description of the most basic term in the model would undermine the potential viability of judging its research value. We claim that the definition—*a* thing *is what can be created, processed, received, released, and transferred*—is of some worth in making the term more well-defined and less fuzzy. Malfouris [39] explains:

The notion of thinging seeks to encapsulate the major phenomenological ingredients . . . , shifting our attention away from the sphere of isolated and fixed categories (objects, artefacts, etc.) to the sphere of the fluid and relational transactions . . . [39]

- Current approaches to things are somewhat limited in comparison to TM. Heidegger [5] emphasized only the ontological thinging of a thing (*producing* [5] – *creation* in the TM) in response to "what is – ness". Heidegger's "thing" is the name we give to a discrete yet unspecifiable entity [7].

- The TM's definition of thing broadens its characterization by including other secondary aspects: process-ness, receiveness, transfer-ness, and release-ness. All four features form possible "thingy Qualities" [3] after production (creation). In a TM, "things" take the characteristics of "objects" as discrete specifiable entities.

A thing that has been created refers to a thing that has been born, is acknowledged, exists, appears, and emerges as a separate item in reality and with respect to other things. A black swan was acknowledged as a metaphor based on a pre-1697 observation that all swans are white. In this case, a black swan was created as a metaphorical thing before 1697. This metaphor was processed and communicated among people at that time. It is the black swan machine. In 1697, a black swan was created in the sphere of knowledge by the appearance of the physical thing. The black swan machine is now a machine that consists of the metaphor and the bird. Note that a thing is a machine and vice versa. A factory can be a thing that is constructed and inspected as well as a machine that receives other things (e.g., materials) to create products. A factory is a thing when it processed (e.g., created), and it is a machine when it is processing things (e.g., creating products).

Is there a thing in a machine—or world—that is not created? Here, creation may refer to physical things (e.g., the sky or an animal), social things (e.g., a society or a celebration), mental things, (e.g., a thought, a feeling, or literature), and nonphysical things (e.g., music). Note that some machines are only processors, receivers, releasers, and/or transferors of a thing. Thus, processing, receiving, releasing, and transferring are important in defining a thing in a noncreating machine.

It is clear why we have opted to use the term *thing* instead of *object*, which, in Heidegger's [5] view, is a manufactured thing, such as a computational *artifact* (e.g., computer-oriented, manufactured data). A thing can be created, processed, received, released, and/or transferred.

*Create* a thing means that it *comes about* and this implies the possibility of its un-thinging within a machine. A collection of machines of a thing forms a larger machine. The stomach machine is a food-processing machine in the digestive machine. The digestive system is one machine in the human being machine, with respect to the thing, food, which is digested (processed) to create waste. A human being is a thing in a school machine.

*Processing* indicates a type of change that a machine performs on a thing without turning it into a new thing (e.g., a car is processed when its color is changed).

*Receiving* is the flow of a thing to a machine from an outside machine. *Releasing* is exporting a thing outside the machine. It stays as a released thing if the exporting channel is not available. *Transferring* is the released thing departing to outside the machine.

Note that the relevant purpose involves things/machines that are relevant for this purpose. After all, a machine is a thing. For example, in a hospital, a human being includes broken or nonfunctioning machines or infectious machines (viral or bacterial machines), and other characteristics used to represent a human thing (machine).

The world of a TM consists of an arrangement of machines, wherein each thing has its own unique stream of flow. TM modeling puts together all of the things/machines required to assemble a system (a grand machine).

The example below illustrates these concepts in terms of the software engineering sphere.

### III. EXAMPLE

Deitel and Deitel's book *C++ How to Program* [40] gives an object-oriented program that uses the class *Time*:

Functions:
    void setTime(int, int, int); // set hour, minute, second
    void printUniversal();       // print universal-time format
    void printStandard();        // print standard-time format
The attributes: int hour, int minute, and int second.
The main program includes such statements as:
    Time t;  // instantiate object t of class Time
    t.printUniversal();  // 00:00:00
    t.printStandard();   // 12:00:00 AM
    t.setTime(13, 27, 6);  // change time

Fig. 3 shows the TM's static (independent of time) representation of this program. Note that in the following discussion Time denotes the class *Time*, while time denotes the notion of time as generally understood and used.

In the figure, the Time machine (circle 1) includes the hour (2), minute (3), and second (4) submachines. All Time submachines are fed by the integer machine (5). When an integer is created, it is processed to verify its constraints (e.g., the second must be between 0 and 60). The created *Time thing* (instance) is processed (6 and 7) in the Time machine to either convert it to the standard or universal format, after which it will flow (8) to the printer to be printed (9).

Fig. 3 represents the program's static description, which we call the *machine*. To specify the *behavior* of the C++ program's execution, we identify different possible events and their chronologies.

An event is a machine in a TM that contains at least three submachines: the time, the region, and the event itself. The region is where the event *takes place* or site of its unfolding. We can bring here Heidegger's notion of gathering, in the sense that the event brings into presence the value (meaningfulness) of the region that was previously hidden. Thus, the event (as a machine) emerges as a thing by gathering (enclosing) the time and region (and other things). Such dwelling (Heidegger's term) can be applied to all phases of the TM modeling, but we want to emphasize engineering here, not philosophical thought.

Fig. 4 shows the representation of the event: *Create the constructor of the class Time*. It includes the three machines: the region of the event (circle 1), which is a subdiagram of Fig. 3; the (real) time submachine (2); and the event submachine itself (3).

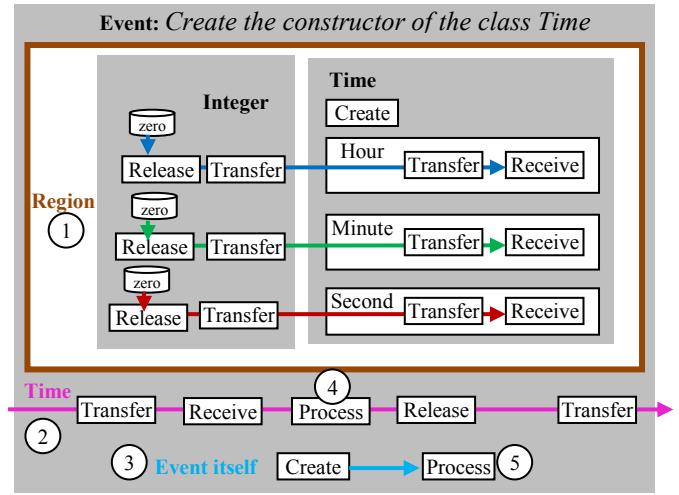

Figure 4. The event creating the constructor.

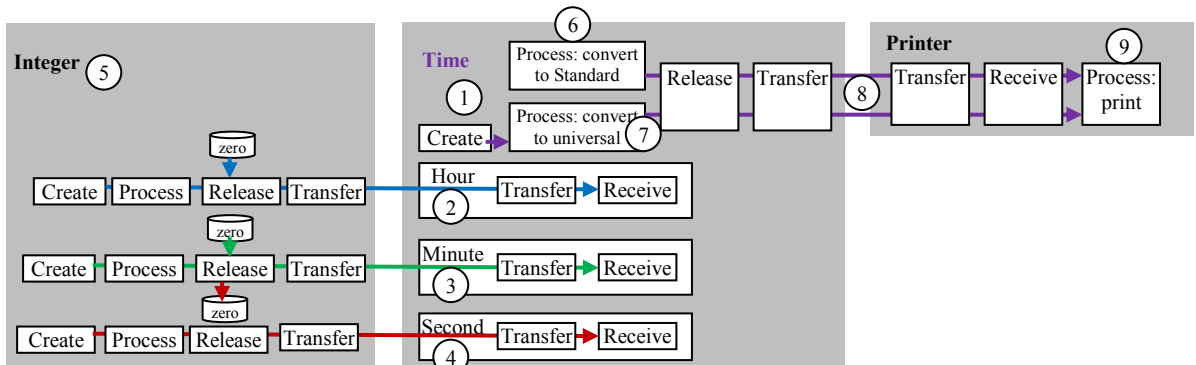

Figure 3. The TM representation of the class Time.

Note that, in general, an event may have other features, such as its intensity. In the figure, the processing of time (4) reflects the consumption of time, whereas the processing of the event (5) indicates that the event is taking its course. For the sake of simplification, we will represent an event only by its region.

Accordingly, we identify the following four events:

Event 1 (E₁): *Create the constructor of the class Time* (Fig. 4);

Event 2 (E₂): *Set Time* (Fig. 5);

Event 3 (E₃): *Print Time in standard form* (Fig. 6); and

Event 4 (E₄): *Print Time in universal form* (Fig. 7).

Fig. 8 shows the chronology of execution of these events. Fig. 9 represents the execution of Deitel and Deitel's program [40].

In philosophical language, this chronology of events is an ordering setting-up, through which "enframing" (gathering things/machines together) is applied to all submachines to enable the program machine to "reveal" itself.

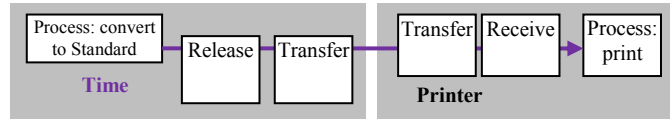

Figure 7. The event *Print Time in universal form*.

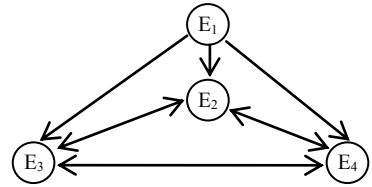

Figure 8. Chronology of the execution.

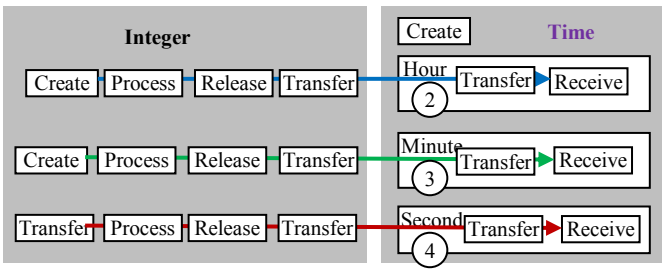

Figure 5. The setting of *Time*.

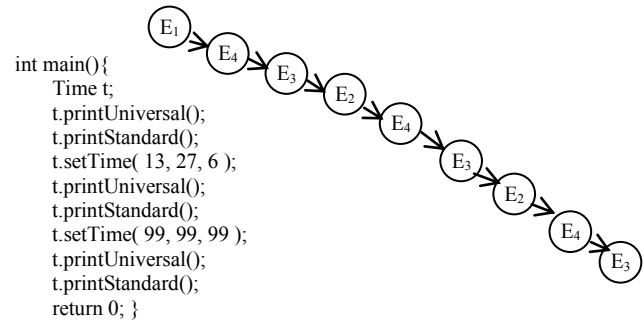

```
int main(){
    Time t;
    t.printUniversal();
    t.printStandard();
    t.setTime( 13, 27, 6 );
    t.printUniversal();
    t.printStandard();
    t.setTime( 99, 99, 99 );
    t.printUniversal();
    t.printStandard();
    return 0; }
```

Figure 9. A sample C++ program and its events.

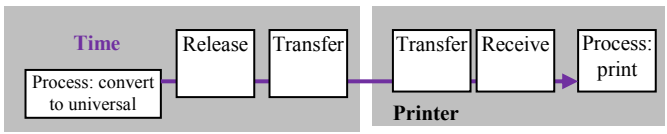

Figure 6. The event *Print Time in standard form*.

## IV. PROCESS VERSUS MACHINE

What is the difference between a machine and a *PROCESS*? The term *PROCESS*, written in capital letters to avoid confusion with Process (change) in the TM machine, denotes what is typically defined as a sequence of operations that transforms input into output. According to Tanaka [41], PROCESS is "a collection of steps taking place in a prescribed manner and leading to an objective."

### A. An Example from the PROCESS Specification Language

In the PROCESS specification language (PSL), a standard exchange language for PROCESSing information in the manufacturing industry [42-43]. "Most PROCESS models support the notions of input and output, which are data or objects provided to a behavior execution before it starts, and data produced when it finishes, respectively" [44].

Bock and Gruninger [44] use Fig. 10 to show an example of a PROCESS change in a car's color using one of the UML 2 notations for object flow. According to Bock and Gruninger [44],

[The figure] is ambiguous not because it is graphical, textual languages have the same problem, but because it is specifying execution with constructs that only implicitly refer to runtime, rather than explicitly. For example, the nodes labeled ChangeColor, Paint, and Dry will be executed many times in many situations, and the diagram does not clarify which executions are referred to, or how the graphical nesting and arcs constrain them.

In addition, Bock and Gruninger [44] use Fig. 11, called the occurrence tree, to demonstrate a runtime execution of an activity: "It has no analog in UML, because UML does not have a direct model of runtime execution yet" [44].

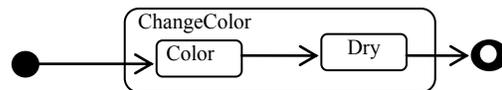

Figure 10. Example UML 2 (redrawn from Bock and Gruninger [44]).

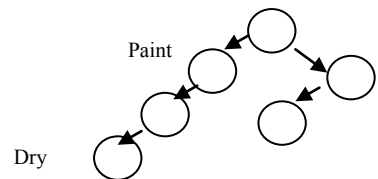

Figure 11. PSL occurrence tree (partially redrawn from Bock and Gruninger [44]).

Fig. 12 shows the corresponding TM representation. Note that the coloring/drying machine includes the PROCESSes of transferring, receiving, releasing, and processing (change). This TM representation, which is illustrated in Fig. 12, can be used to specify the execution of a sequence of events. Fig. 13 shows two possible *thingings* of events; each of them represent a different "slicing" of regions in Fig. 12, depending on the design mode of *thinging* for the events.

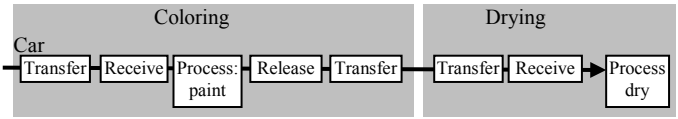

Figure 12. TM representation of the color/dry machine.

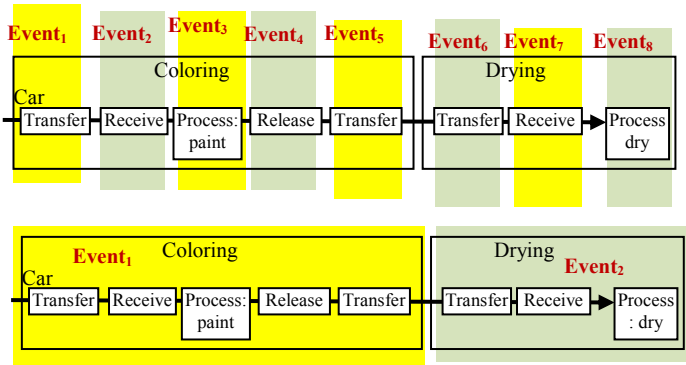

Figure 13. Two possible thingings of events.

## B. Chronology of Events

To make the example more compelling, let us assume that
- A first test is performed to check whether the car has been colored to a satisfactory level, and
- A second test is conducted to check whether the car is completely dry.

Accordingly, Fig. 14 shows the new TM representation. We inserted the machine testing after the car is first colored (circle 1 in the figure). If the paint is satisfactory, the car continues to drying (2); otherwise, the car is sent back (3) to be painted again (4).

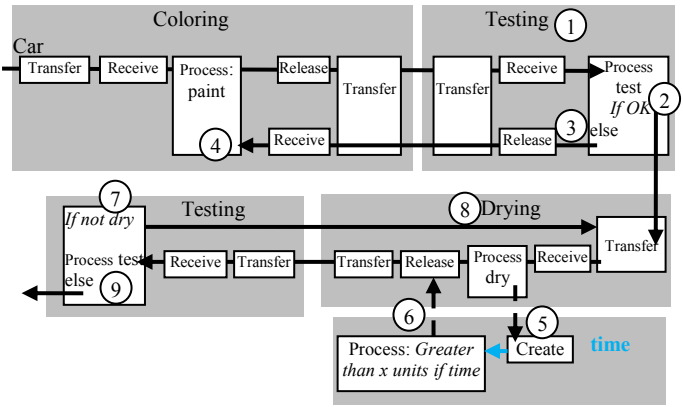

Figure 14. The TM representation with testing.

When drying starts, a time is set (5) (e.g., 1 hour). At the end of that time, the car is checked (6). If it is not dry, the car is dried again for the set time. If the car is dry, then it has been finished (9).

To model the machine's behavior for a single car, Fig. 15 shows seven selected events:

Event 1 ($E_1$): A car arrives and is painted.
Event 2 ($E_2$): The car is tested to see whether the paint is satisfactory.
Event 3 ($E_3$): The car is returned to be repainted.
Event 4 ($E_4$): The car is dried.
Event 5 ($E_5$): The car is sent to be tested for dryness.
Event 6 ($E_6$): The car is sent back to be dried again.
Event 7 ($E_7$): The car is dry and released from the station.

Fig. 16 renders the chronology of these events, exemplified by a single car.

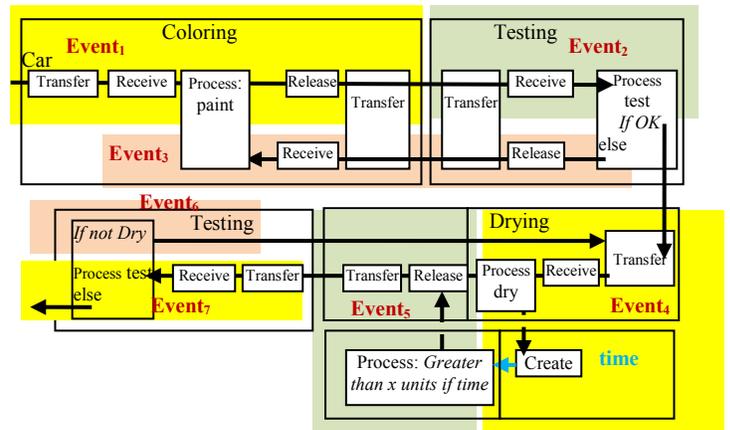

Figure 15. Events.

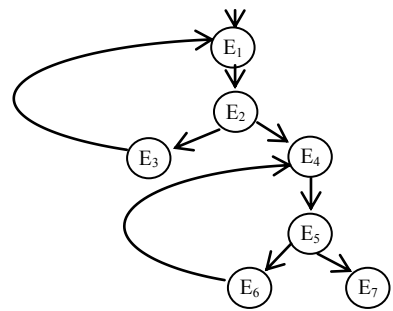

Figure 16. The chronology of execution for a car.

Note that the thinging of the events is rendered pursuant to their "meaningfulness" to the modeler of the coloring/drying machine. In $E_1$, *a car arrives and is colored*. Subevents such as receiving the car are not of interest (e.g., to report, register, note these events), so they are subsumed in $E_1$. This is an example of thinging events. The execution of events in Fig. 15 represents the lifecycle of a single car in the coloring/drying machine (one *occurrence* of the behavior of the machine). It is interesting to investigate the intersection of multiple occurrences.

## C. Behavior with Multiple Cars

Consider a situation in which we can maximize the use of the coloring/drying machine with multiple cars. In this case, we have to add queues to the coloring and drying submachines, as shown in Fig. 17. In Fig. 17, cars arriving in the coloring machine are queued (circle 1). They are processed one by one (2). When a car is being colored, the state of the coloring submachine is set to *busy* (3). When a car leaves the coloring submachine, the state is set to *not busy* to allow another car from the queue to be colored. A similar procedure is installed in the drying machine (5, 6, 7, and 8).

Fig. 18 shows the results of thinging "meaningful" events, whereas Fig. 19 shows the chronology of the events for one car. To "run" (execute) the coloring/drying machine such that multiple cars can be processed simultaneously, we simplify the process by assuming that no car is returned to be colored or dried twice; that is, the color and dryness are satisfactory after the first time. Fig. 20 demonstrates a situation with different cars during which events overlap. Car 1 "enters" E$_1$ and then flows to E$_2$. As soon as it "leaves" from E$_2$ to E$_3$, car 2 "enters" E$_2$. Accordingly, different cars progress to different events of the coloring/drying machine. Eventually, in the last column of Fig. 20, seven cars are being processed simultaneously.

This illustrates parallel car processing chronologies (multiple iterations of Fig. 19) in the behavior of the machine. This illustration of the TM's specification advances smoothly from thinging things and machines to thinging events to modeling the machine's dynamic activity.

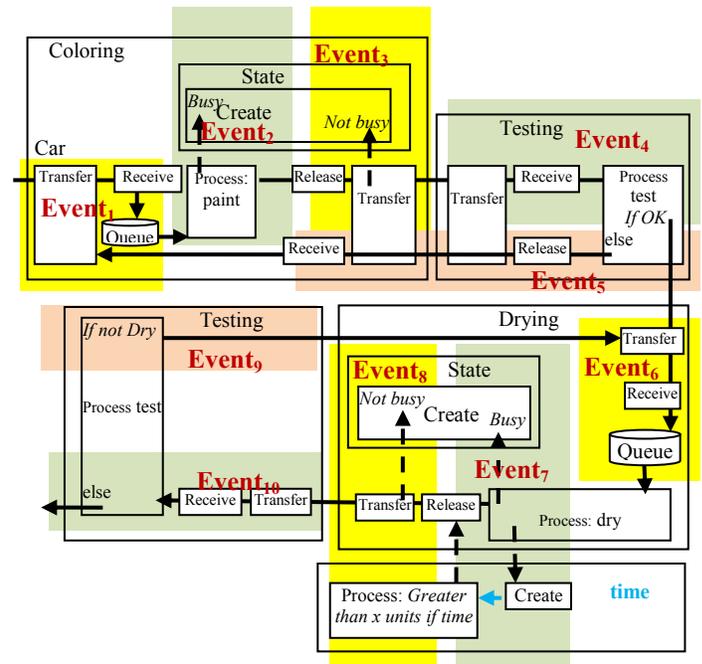

Figure 18. Events in the example.

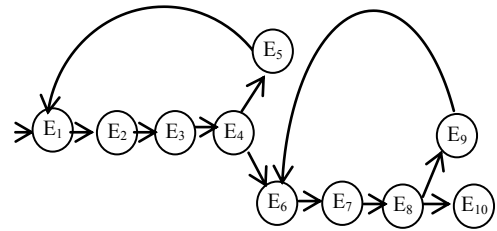

Figure 19. The chronology of execution for a car using the queue.

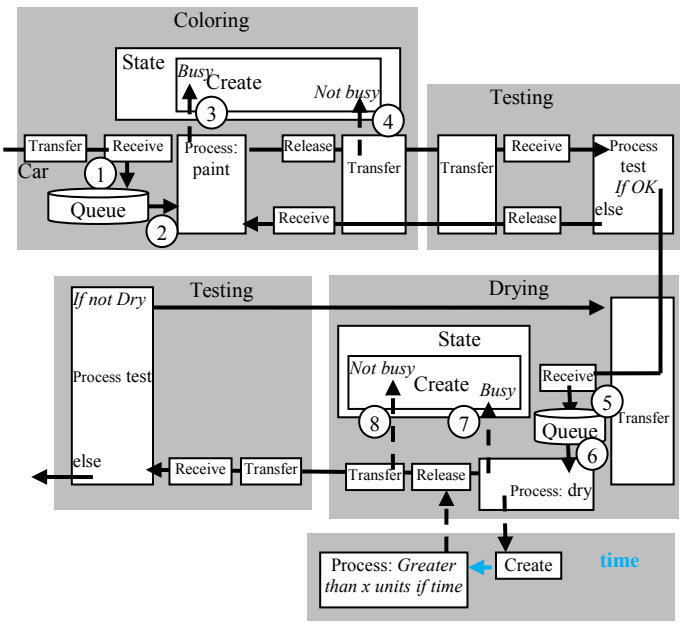

Figure 17. The TM representation with queues.

| Sequence of arrivals | Car 1 | Car 2 | Car 3 | Car 4 | Car 5 | Car 6 | Car 7 |
|---|---|---|---|---|---|---|---|
| | E$_1$ | E$_1$ | E$_1$ | E$_1$ | E$_1$ | E$_1$ | E$_1$ |
| Time period 1 | E$_2$ | | | | | | |
| Time period 2 | E$_3$ | E$_2$ | | | | | |
| Time period 3 | E$_4$ | E$_3$ | E$_2$ | | | | |
| Time period 4 | E$_6$ | E$_4$ | E$_3$ | E$_2$ | | | |
| Time period 5 | E$_7$ | E$_6$ | E$_4$ | E$_3$ | E$_2$ | | |
| Time period 6 | E$_8$ | E$_7$ | E$_6$ | E$_4$ | E$_3$ | E$_2$ | |
| Time period 7 | E$_{10}$ | E$_8$ | E$_7$ | E$_6$ | E$_4$ | E$_3$ | E$_2$ |

Figure 20. The chronology of execution for multiple cars.

## V. THINGING AND EVENTING

The previous section stated that an event in the TM includes at least three submachines: the time, region, and event itself. According to Heidegger [5], "the particularity of things seems to depend completely on their *space* and *time*" [44]. Space in the TM is called the region of the event, as shown in the previous examples. From this perspective, TM events are of sources that generate particularities. If we focus on the space aspect of regions in events, we find that it is a logical space:

> Even if we break a thing to get to the space "inside" we find external relations between its parts, bits, and pieces. Space seems to be not really "in" the thing but only the "possibility" of arrangements of its parts (in, out, next to, etc.). [44]

For Heidegger [5], time and space are the realms in which things can be given. Edwards [33] expands, "They stabilize the flow of sentience; they make it into something. They bring it to a lasting stand". In TM events, time and space even out the flow of things (e.g., in the Time class example, the three flows of integers [hour, minute, and second] reach their destination to create a particular time [thing]). Analogous to analyzing a connection between a subject and a predicate [44], the TM conceptualizes a connection between a machine (system) and an event (region/time diagram). In the context of the machine, the region diagram expresses itself as a situation in which facets of itself are stated and in which something (the creation) is asserted about the thing. (The last two sentences express an alternative account of Gendlin's [44] description of a connection between a thing and a human being.)

## VI. CONCLUSION

This paper has presented the term *thing* and showed that this specific term and the general term *thinging* are valued notions that play an important role in the need to distinguish separable entities in software engineering modeling. Additionally, the paper attempted to answer the question, what is a thing, object, or process? As a result, this may raise the issue of thinging in software engineering.

An (abstract) TM is reintroduced and proposed as a foundation for the clarification of these notions.

The substance of this paper suggests that thinging should be more meaningfully emphasized as a research and teaching topic, at least in the requirement analysis phase of the software development cycle. According to Umans [45], "the quest for knowledge is not a quest for truth but a challenge to understand the processes of thinging."